\documentclass[aps,prl,twocolumn,superscriptaddress,a4paper,floatfix]{revtex4-2}
\usepackage[utf8]{inputenc}
\usepackage{graphicx}
\usepackage{amsmath,amssymb}
\usepackage[colorlinks=true, linkcolor=black, citecolor=black, urlcolor=blue]{hyperref}
\usepackage{datetime}
\usepackage{siunitx}
\usepackage{braket}
\usepackage{cleveref}
\usepackage{gensymb}
\usepackage{mathtools}
\usepackage{dsfont}
\usepackage{braket}
\usepackage{comment}
\usepackage{physics}
\usepackage{xcolor}
\usepackage{fdsymbol}
\usepackage{tikz}

\definecolor{devon9_1}{HTML}{2b194c} 
\definecolor{devon9_2}{HTML}{293367}
\definecolor{devon9_3}{HTML}{275084}

\definecolor{devon9_4}{HTML}{3569ad} 
\definecolor{devon9_5}{HTML}{6181d1}
\definecolor{devon9_6}{HTML}{989be7}

\definecolor{devon9_7}{HTML}{bab4f1} 
\definecolor{devon9_8}{HTML}{d1cdf6}
\definecolor{devon9_9}{HTML}{e8e6fb}

\definecolor{batlow9_1}{HTML}{011959}
\definecolor{batlow9_2}{HTML}{103f5f}
\definecolor{batlow9_3}{HTML}{1a5862}

\definecolor{batlow9_4}{HTML}{396e58}
\definecolor{batlow9_5}{HTML}{647e43}
\definecolor{batlow9_6}{HTML}{9a9132}

\definecolor{batlow9_7}{HTML}{d0a35a}
\definecolor{batlow9_8}{HTML}{efb094}
\definecolor{batlow9_9}{HTML}{ffd8d6}

\definecolor{batlow12_1}{HTML}{011959}
\definecolor{batlow12_2}{HTML}{0e375e}
\definecolor{batlow12_3}{HTML}{134c61}

\definecolor{batlow12_4}{HTML}{205f61}
\definecolor{batlow12_5}{HTML}{396e58}
\definecolor{batlow12_6}{HTML}{597a48}

\definecolor{batlow12_7}{HTML}{7e8737}
\definecolor{batlow12_8}{HTML}{a99635}
\definecolor{batlow12_9}{HTML}{d0a35a}

\definecolor{batlow12_10}{HTML}{e9ac86}
\definecolor{batlow12_11}{HTML}{fac0b5}
\definecolor{batlow12_12}{HTML}{ffe4e3}

\definecolor{bilbao9_1}{HTML}{4c0000}
\definecolor{bilbao9_2}{HTML}{732429}
\definecolor{bilbao9_3}{HTML}{93444a}

\definecolor{bilbao9_4}{HTML}{a06257}
\definecolor{bilbao9_5}{HTML}{a6775c}
\definecolor{bilbao9_6}{HTML}{ac8d60}

\definecolor{bilbao9_7}{HTML}{b5a772}
\definecolor{bilbao9_8}{HTML}{c3bda4}
\definecolor{bilbao9_9}{HTML}{d3d2ce}

\DeclareRobustCommand\sampleline[1]{%
  \tikz\draw[#1, line width=1.8pt] (0,0) (0,\the\dimexpr\fontdimen22\textfont2\relax)
  -- (1.9em,\the\dimexpr\fontdimen22\textfont2\relax);%
}

\newcommand{\EPFL}{Institute of Physics and Center for Quantum Science and Engineering, Ecole Polytechnique F\'ed\'erale de Lausanne (EPFL), CH-1015 Lausanne, Switzerland}
\newcommand{\ETH}{Institute for Theoretical Physics, ETH Zürich, CH-8093 Zürich, Switzerland}
\newcommand{\Harvard}{Lyman Laboratory, Department of Physics, Harvard University, Cambridge, MA 02138, USA}
\begin{document}

\title{Cavity-mediated charge and pair-density waves in a unitary Fermi gas}

%
%
%
%


\author{T.~Zwettler}
\affiliation{\EPFL}
\author{F.~Marijanovic}
\affiliation{\ETH}
\author{T.~Bühler}
\affiliation{\EPFL}
\author{S.~Chattopadhyay}
\affiliation{\ETH}
\affiliation{\Harvard}
\author{G.~Del Pace}\altaffiliation{Current affiliation:Department of Physics, University of Florence, INO-CNR and European Laboratory for Nonlinear Spectroscopy (LENS), University of Florence, 50019 Sesto Fiorentino, Italy}
\affiliation{\EPFL}
\author{L.~Skolc}
\affiliation{\ETH}
\author{V.~Helson}
\affiliation{\EPFL}
\author{S.~Uchino}
\affiliation{Faculty of Science and Engineering, Waseda University, Tokyo 169-8555, Japan}
\author{E.~Demler}
\affiliation{\ETH}
\author{J.P.~Brantut}
\affiliation{\EPFL}

\date{\today}

\begin{abstract}
Coherent light-matter interactions between a quantum gas and light in a high-finesse cavity can drive self-ordering phase transitions. To date, such phenomena have involved exclusively single-atom coupling to light, resulting in coupled charge-density or spin-density wave and superradiant order. In this work, we engineer simultaneous coupling of cavity photons to both single atoms and fermionic pairs, which are also mutually coupled due to strong correlations in the unitary Fermi gas. This interplay gives rise to an interference between the charge-density wave and a pair-density wave, where the short-range pair correlation function is spontaneously modulated in space. We observe this effect by tracking the onset of superradiance as the photon-pair coupling is varied in strength and sign, revealing constructive or destructive interference of the three orders with a coupling mediated by strong light-matter and atom-atom interactions. Our observations are compared with mean-field theory where the coupling strength between atomic- and pair-density waves is controlled by higher-order correlations in the Fermi gas. These results demonstrate the potential of cavity quantum electrodynamics to produce and observe exotic orders in strongly correlated matter, paving the way for the quantum simulation of complex quantum matter using ultracold atoms.
\end{abstract}
\maketitle


The interplay of competing and cooperating emergent orders is a hallmark of strongly correlated quantum matter \cite{Sachdev2012,fernandes_intertwined_2019}. These systems --ranging from high-temperature superconductors \cite{RevModPhys.87.457,ZAANEN1999217,PhysRevB.63.094503} to Van der Waals materials \cite{doi:10.1126/science.abm8386,Balents2020,Nandkishore2012} and multiferroics \cite{Wang01072009,fiebig_evolution_2016,KHOMSKII20061} -- exhibit diverse phases within narrow parameter ranges, offering remarkable tunability and technological promise. Despite intense research efforts, understanding these coupled orders remains a significant challenge, as they arise from complex quantum correlations that defy simple conceptualization. Quantum gas experiments provide a powerful platform for exploring complex ordering phenomena, as they combine a microscopic Hamiltonian that is known \textit{a priori}, interactions which can be tuned to extreme regimes, and direct observation of the different macroscopic orders. This makes quantum gases an ideal environment for unraveling the mechanisms driving emergent order in strongly correlated materials. In this context, cavity-quantum electrodynamics (cQED) methods are particularly suitable for studying the interplay of distinct emergent orders in a quantum system \cite{mivehvar:2021aa}. Indeed, tunable, coherent light-matter interactions can be created and combined with the short-range interactions in a quantum gas, resulting in the emergence of superradiant phases \cite{Black:2003aa,Slama:2007ab,Baumann:2010aa,Klinder:2015aa,zhang:2021tr} that coexist with strong contact interactions \cite{Klinder:2015ab,Landig:2016aa,HelsonDWOIAUFGWPMI2023}. 

In this work, we leverage the versatility and control of cQED in a many-body setting to experimentally realize the interplay of three strongly coupled quantum orders in a quantum-degenerate Fermi gas within a high-finesse optical cavity: photonic superradiance, charge-density wave and pair-density wave. Our experiment operates in the unitary limit of contact interactions, where fermions form pairs with a size on the order of the interparticle spacing \cite{randeria:2014aa}. The cavity light field dispersively couples to single atoms and atomic pairs simultaneously \cite{KonishiUPPIASIFG2021}. In the presence of a transverse pump, this dual coupling allows for the formation of both charge-density wave and pair-density wave order, which are also mutually coupled via strong and tunable atom-atom interactions by a Feshbach resonance. Light scattered off single atoms and atomic pairs interferes, giving rise to a characteristic Fano-type profile of the superradiant phase diagram. This profile reflects the competition and cooperation between the two fermionic orders. On the competitive side the suppression of the onset of the superradiant phase can be understood as frustration between antagonistic density wave and pair-density wave orders. A mean-field analysis of the three coupled order parameters captures this effect. By employing a set of linear-response functions, our analysis reveals the role of previously unexplored higher-order correlations of the Fermi gas, offering insights into the complex behavior of multiple competing and cooperating coupled orders.

\begin{figure}[h!]
\includegraphics{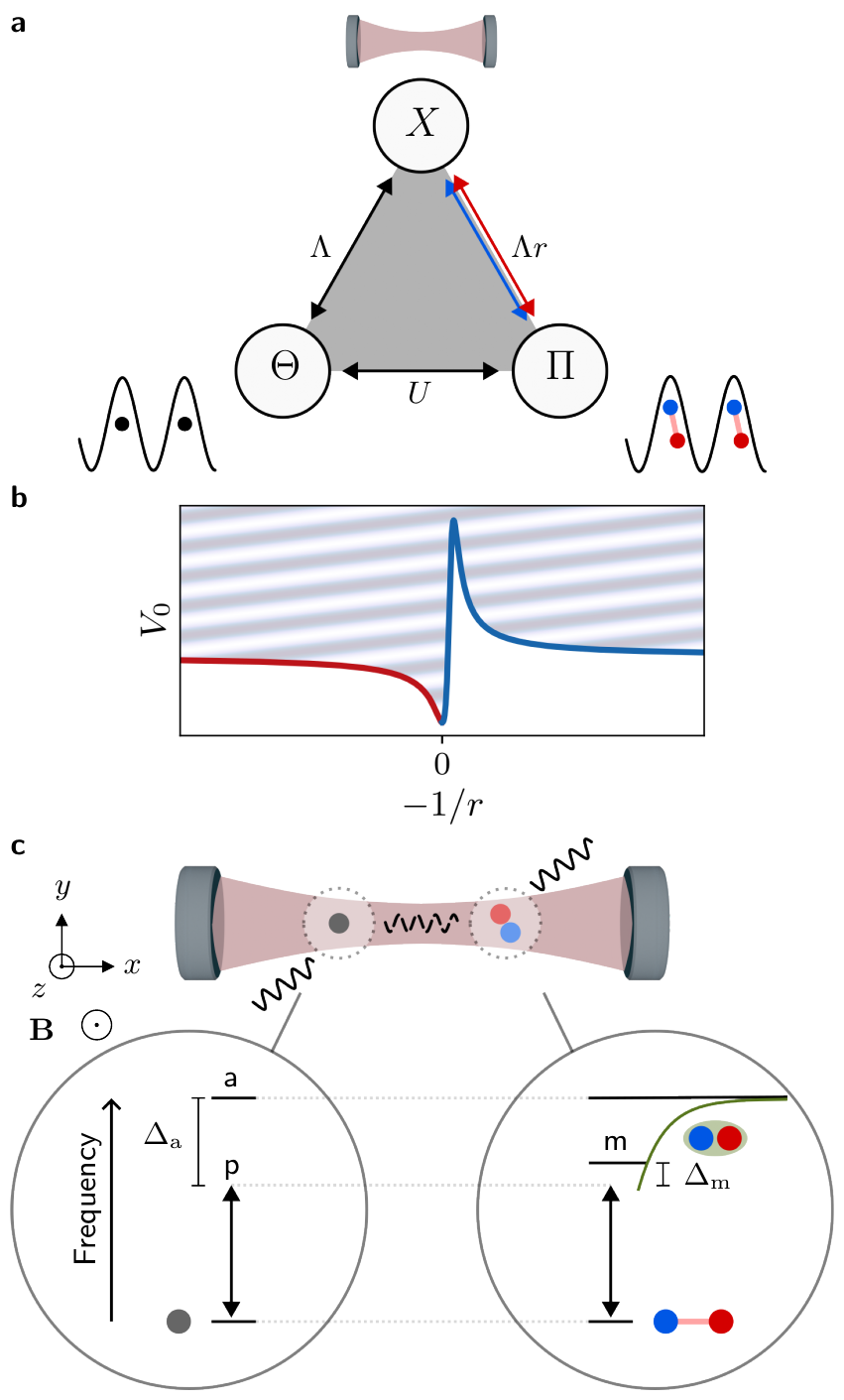}
\caption{\textbf{Coupled order parameters.} {\bf a} Charge-density wave ($\Theta$), pair-density wave ($\Pi$), and in-phase cavity-field quadrature ($X$) represent the three coupled orders. $\Lambda$, the strength of the dispersive atom-cavity coupling, is proportional to the pump strength $V_0$. $r$ denotes the relative strength of the dispersive coupling of atoms and pairs to the cavity. 
$\Theta$ and $\Pi$ are mutually coupled by strong atom-atom interactions with a strength $U$. {\bf b} Schematic phase boundary separating the normal ($X=0$) from the superradiant phase ($X\neq0$) as a function of $V_0$ and $-1/r$, showing a characteristic Fano-type profile. {\bf c} Single atoms (left) and pairs of atoms (right) in a unitary Fermi gas within the mode of a high-finesse cavity can scatter photons from a transverse pump into the cavity and vice-versa. The dispersive coupling strengths $\Lambda$ and $r\Lambda$ are determined by the detuning $\Delta_\mathrm{a}$ between the pump (p) and the atomic resonance (a), and the detuning $\Delta_\mathrm{m}$ between the pump and a photo-association resonance (m), respectively.}
\label{fig:fig1}
\end{figure}

\section{Coupled order parameters}

We realize a system with three strongly coupled order parameters, describing the in-phase quadrature of the cavity field and the amplitudes of the charge and pair-density waves, described by the macroscopic fields $X$, $\Theta$, and $\Pi$,  respectively. Up to second order in the macroscopic fields, the Landau-Ginzburg free energy of the system can be written as:
\begin{equation}
    \begin{split}
        \mathcal{F}(\Theta,\Pi,X) =  &\epsilon_{\Theta} \Theta^2 + \epsilon_{\Pi} \Pi^2 + \epsilon_X X^2 \\
        & - \rm{U} \Theta \Pi - \Lambda X(\Theta + r \Pi)
        \label{eq:freeEnergy}
    \end{split}
\end{equation}
and schematically represented as in Fig.~\ref{fig:fig1}a (see Method for a microscopic derivation). $\epsilon_\Theta, \epsilon_\Pi$ and $\epsilon_X$ represent the energy cost of the uncoupled orders, which are all positive, and therefore oppose ordering. The coupling between $\Theta$ and $\Pi$ is set by $U$, the coupling between $X$ and $\Theta$ is determined by $\Lambda$, and that between $X$ and $\Pi$ by $r \Lambda$. Ordering occurs when, due to their mutual coupling, the curvature of the free energy in the vicinity of the origin is negative along a particular direction. In our experiment we fix $U$ and vary $\Lambda$ and $r$, tuning the latter in both sign and strength. Choosing the same sign for the three couplings results in intertwining, where all the ordering contributions cooperate. Conversely, flipping the sign of $r$ leads to frustration and competing order parameters. The phase diagram for this system is sketched in Fig. \ref{fig:fig1}b, with a boundary separating the organized and homgeneous phases exhibiting a characteristic Fano shape, as a function of $1/r$. 
Akin to geometric frustration, this phenomenology is only possible with more than two coupled orders. With only two orders such as light and atomic density, the sign of the coupling is irrelevant as it amounts to a sign change of the pump beam amplitude.  

Microscopically, the simultaneous occurrence of strong atom-atom and strong light-matter interactions couples the three orders as follows (see Methods for the formal derivation). First, the parameter $\Lambda$ has its strength and sign controlled by the detuning and power of a pump laser beam illuminating the atoms from the side, as depicted in Fig.~\ref{fig:fig1}c. It arises from Rayleigh scattering of photons from the pump into the cavity mode by the atomic gas. Second, the pair and charge-density waves are intrinsically coupled by the contact interactions in the unitary Fermi gas. Manifestations of this coupling can be found, for example, in the equation of state, where the two thermodynamic quantities describing density and pair-density have mutual dependence. Third, our system also features a direct, dispersive coupling between photons and fermion pairs with strength $r\Lambda$, by operating the cavity resonance at a finite frequency difference $\Delta_m \propto -1/r$ of a photoassocation (PA) transition \cite{KonishiUPPIASIFG2021}, as shown in Fig.~\ref{fig:fig1}c. The parameter $r$ can be interpreted as quantifying the interference between photons scattered off atoms on the one hand, and pairs of atoms on the other hand, into the cavity. Importantly, the direct coupling between the density and the pair-density orders contrasts with previous realizations of coupled orders in the cavity QED context, where different atomic modes scatter photons between a set of optical modes \cite{Morales:2018aa}, or in the case of magnetic textures \cite{Kroeze:2018aa,Landini:2018aa,kongkhambut:2021aa} where cross-couplings occur due to dissipation \cite{Dogra:2019aa}.

\begin{figure}[h!]
\includegraphics{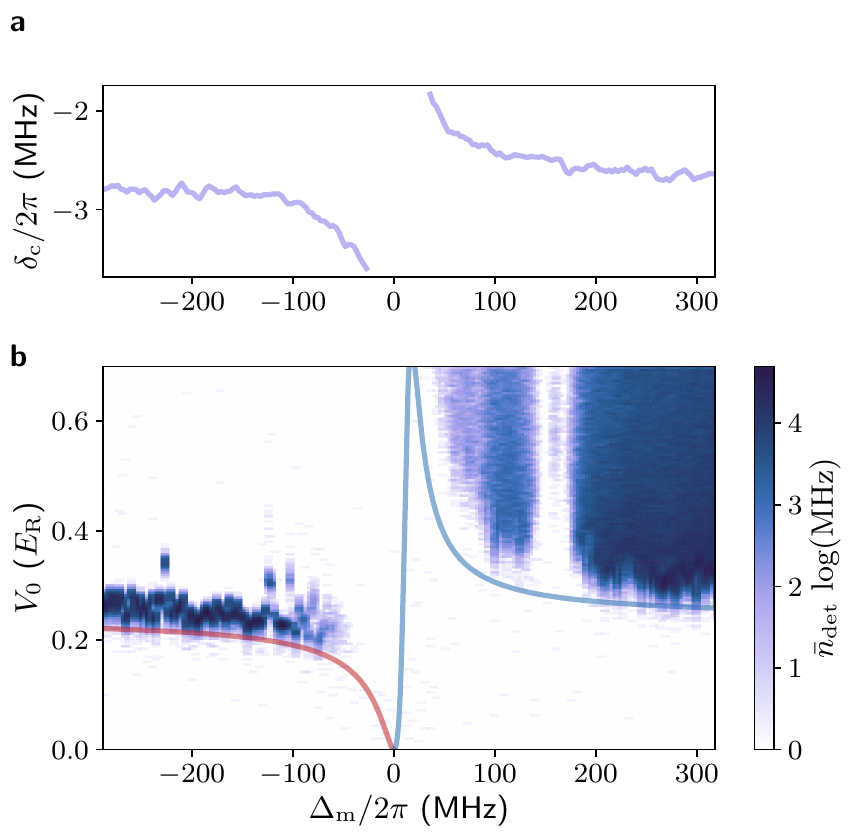}
\caption{\textbf{Ordering close to a photo-association transition.}{\bf a} Dispersive shift $\delta_\mathrm{c}$ measured by transmission spectroscopy as a function of the molecular detuning $\Delta_\mathrm{m}$, showing the avoided crossing pattern characteristic of strong light-matter coupling.  {\bf b}  Photon flux traces as a function of pump strength $V_0$ while varying $\Delta_\mathrm{m}$ across the photoassociation transition at fixed $\Delta_\mathrm{c} = -5.5~\mathrm{MHz}$. 
The blue-red line indicates the Fano-type phase boundary predicted in Fig.~\ref{fig:fig1}b, with an overall scaling factor left as a free parameter for the $V_0$ axis, and the other parameters calculated using a mean-field theory.}
\label{fig:fig2}
\end{figure}

\section{Experimental observation}

We investigate ordering through the observation of the onset of superradiance, as the relative coupling strength between light and atoms, and light and pairs, is increased. The experiment starts with a quantum degenerate, unitary Fermi gas comprising $N = 5.2 \times 10^5$ $^6$Li atoms, equally populating the two lowest hyperfine states within the mode of a high finesse cavity (see \cite{buhler:2024aa} and Methods). As depicted in Fig.~\ref{fig:fig2}a, the spectrum of the coupled system close to the PA transition exhibits the characteristic avoided crossing pattern due to strong photon-pair coupling \cite{KonishiUPPIASIFG2021}. We use a retro-reflected, transverse pump beam with an absolute frequency within $\pm 400~\mathrm{MHz}$ of the PA transition. Its intensity is parametrized by the trap depth $V_0$ which it produces at the location of the atoms. It controls the value of $\Lambda$ \cite{HelsonDWOIAUFGWPMI2023}, while the detuning $\Delta_m$ with respect to the PA transition fixes $r$. 

We detune the cavity resonance frequency by $\Delta_{\mathrm{c}} = -5.5$ $\mathrm{MHz}$ from the pump frequency, thereby fixing the parameter $\epsilon_X$. The pump strength $V_0$ is then linearly increased over $\SI{700}{\micro\second}$, and the flux of photons from the cavity $\bar{n}_\mathrm{det}$ is monitored on a single-photon counter, measuring $X^2$ in real time. The transition to the superradiant phase is manifested by a burst of photodetection events, allowing us to locate and track the critical pump strength. Fig.~\ref{fig:fig2}b shows photon flux traces collected as $\Delta_\mathrm{m}$ is varied accross the PA resonance. We observe the superradiant transition for both positive and negative values of $\Delta_\mathrm{m}$, with a pronounced asymmetry, directly demonstrating the contribution of pairs to the signal. Indeed, as $|\Delta_{\mathrm{m}}|$ is reduced, the decrease of the critical pump strength for $\Delta_\mathrm{m} < 0$ and increase for $\Delta_\mathrm{m} > 0$ indicates cooperation and competition between charge and pair-density waves, respectively, and qualitatively reproduces the generic phase diagram presented in Fig.~\ref{fig:fig1}b. Overlayed with Fig.~\ref{fig:fig2}b, we present the phase boundary calculated from a mean-field, linear response theory (see below and Methods), leaving the background $\Delta_\mathrm{m} \longrightarrow \pm\infty$ threshold as a free parameter, showing good agreement and confirming our interpretation in terms of coupled orders. 

In the close vicinity of the molecular transition, the signal is dominated by two-body losses due to spontaneous emission. Additionally, we observe a significant decay of the self-organized phase for $\Delta_\mathrm{m} < 0$ compared to $\Delta_\mathrm{m} > 0$, visible through the absence of cavity photons in the upper-left part of Fig. \ref{fig:fig2}b. We attribute this difference to a combination of a larger photoassociation loss rate at high light intensity on the red side of the PA transition \cite{Chin_Grimm_TiesingaFRIUG}, together with additional optomechanical instabilities due to the larger total dispersive coupling, which are also observed without molecular coupling. Below the critical pump strength however, losses are both low and symmetric between positive and negative $\Delta_\mathrm{m}$ (see Methods and Extended Data Fig. \ref{fig:sup2}), thus the variations in the critical point are directly reflecting the interplay between light-matter and atom-atom interactions. For instance, even though approaching $\Delta_\mathrm{m}=0$ increases losses, the threshold is nevertheless reduced for $\Delta_\mathrm{m}<0$, indicating that the strong dispersive effects of the photon-pair coupling dominates over dissipative mechanisms. The substantial increase in the critical pump strength as $\Delta_{\mathrm{m}} $ approaches zero from the positive side even suggests that a regime where the coupling between $X$ and $\Pi$ is the dominant one. 

\begin{figure}
\includegraphics[width=\columnwidth]{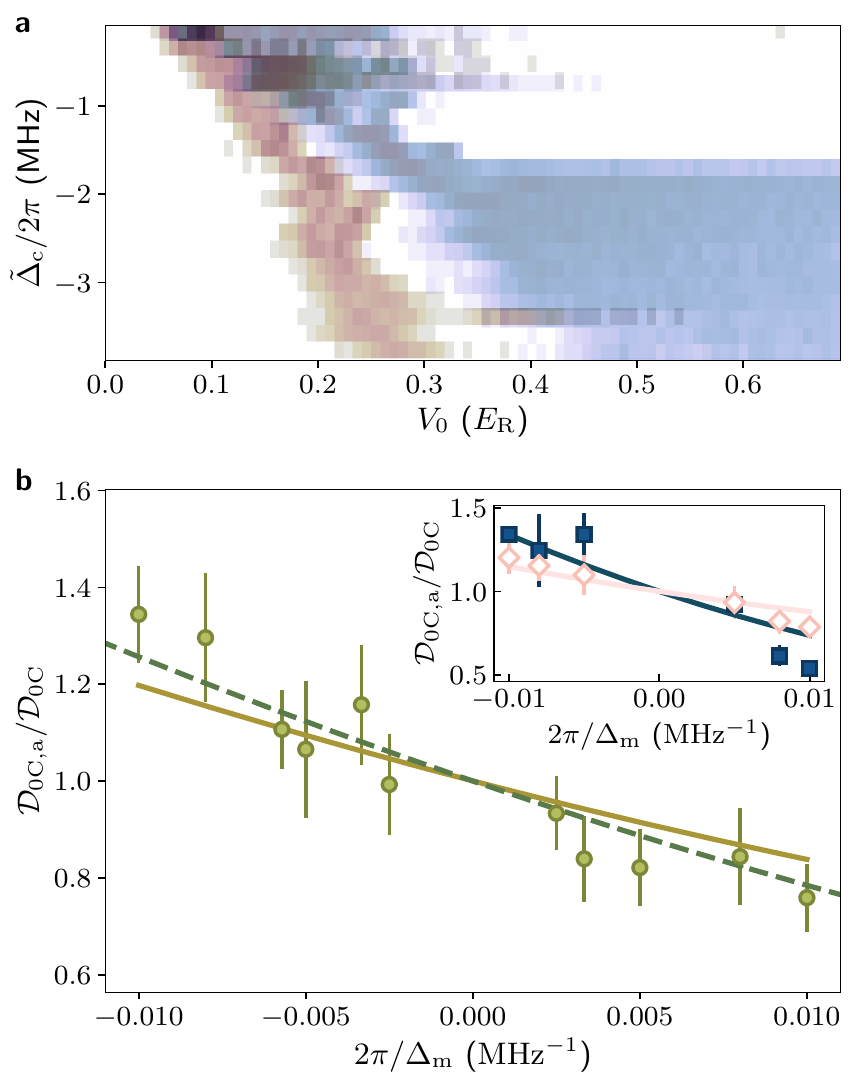}
\caption{{\bf Critical light-matter coupling strength. }{\bf a}. Phase diagrams of density wave ordering as a function of $\tilde{\Delta}_\mathrm{c}$ and pump strength $V_0$, for $\Delta_\mathrm{m} = -100~\mathrm{MHz}$ (red) and $+100~\mathrm{MHz}$ (blue). The growing offset between the two phase boundaries between positive and negative $\Delta_\mathrm{m}$ is a manifestation of the intertwining of density and contact-density wave order. {\bf b}. Normalized critical coupling strength $\mathcal{D}_{0\mathrm{C}, a}/\mathcal{D}_{0\mathrm{C}}$ as a function of the inverse detuning $2\pi/\Delta_\mathrm{m}$ at unitarity ($ \textcolor{batlow12_7}{\medcircle}$) agree with the predictions of the zero-momentum model (\textcolor{batlow12_6}{\sampleline{dashed}}) and the mean-field (\textcolor{batlow12_8}{\sampleline{}}) including trap averaging, within our error bars. The inset depicts measurements away from unitarity at $1/k_\mathrm{F}a = 0.22$ ($ \textcolor{batlow12_2}{\medsquare}$) and $1/k_\mathrm{F}a = -0.16$ ($ \textcolor{batlow12_11}{\meddiamond}$) and the respective mean-field models (\textcolor{batlow12_3}{\sampleline{}}, \textcolor{batlow12_12}{\sampleline{}}).}
\label{fig:fig3}
\end{figure}

To quantitatively connect the changes of the critical pump strength to the atomic and pair-density wave nature of the organized phase, we measured phase diagrams in the $\tilde{\Delta}_\mathrm{c}-V_0$ plane at different $\Delta_\mathrm{m}$, where $\tilde{\Delta}_\mathrm{c}$ is the pump cavity-detuning corrected for the mean dispersive shift. This is illustrated in Fig.~\ref{fig:fig3}a for the example of $\Delta_\mathrm{m} = \pm 100~\mathrm{MHz}$. This approach enables us to separate the direct effect of coupling between orders from the variations in the dispersive shift. Indeed, the pump-cavity detuning is modified by the dispersive coupling to pairs, even in the absence of genuine coupling between orders, making it difficult to quantitatively ascribe the observations of Fig.~\ref{fig:fig2}b to the effect of strong interactions. From each phase diagram, we determine a mean critical reduced light-matter coupling strength $\mathcal{D}_{0\mathrm{C}}$, given by the slope of the linear dependence of the critical pump strengths on the cavity detuning (see Extended Data Fig. \ref{fig:sup1}). In the absence of coupling between light and pairs, this represents the critical strength of the effective interaction between atoms mediated by the cavity and leading to self-organization \cite{ mivehvar:2021aa,HelsonDWOIAUFGWPMI2023}. 

The variations of $\mathcal{D}_{0\mathrm{C}}$ are shown in Fig.~\ref{fig:fig3}b as a function of $1/\Delta_\mathrm{m} \propto -r$, directly quantifying the coupling between the charge and pair density waves. We normalize these values by $\mathcal{D}_{0\mathrm{C,a}}$, corresponding to the atomic contribution alone at $1/\Delta_m =0$. A significant variation of $\mathcal{D}_{0\mathrm{C}}$ is observed, reaching a $35\%$ decrease for cooperative coupling at $1/\Delta_\mathrm{m} < 0$, and a $20\%$ increase for competition at $1/\Delta_\mathrm{m} > 0$, indicating large atom-pair interactions. With a Fermi gas prepared away from unitarity, as illustrated in the inset of Fig.~\ref{fig:fig3}b, the same trend is observed. As we describe below, the larger effect of pair-density waves for positive scattering lengths is due to a stronger cross-coupling between density and pair-density in this regime.

\section{Theoretical interpretation}
    
Having established the correspondence between the experimentally observed Fano lineshape for the superradiant transition and the phenomenology of three coupled order parameters arising from Eq. \eqref{eq:freeEnergy}, we now delve into the microscopics of our system, arguing that indeed the experiment faithfully realizes the physics described by Eq.  \eqref{eq:freeEnergy}. 
The microscopic order parameters for matter can be written as $\Theta = \sum_{Q,\sigma} \int d\mathbf{r
}e^{i\mathbf{Q} \cdot \mathbf{r}} \psi_{\sigma}^\dag(\mathbf{r})\psi_{\sigma}(\mathbf{r})$ and $ \Pi = \sum_{Q} \int d\mathbf{r} e^{i\mathbf{Q} \cdot \mathbf{r}} \psi_{\uparrow}(\mathbf{r}) \psi_{\downarrow}(\mathbf{r}) $, where $\psi_\sigma^\dag(\mathbf{r})(\psi_\sigma(\mathbf{r}))$ is the fermionic creation(annihilation) operator at position $\mathbf{r}$ with spin $\sigma$. $\mathbf{Q}$ runs over the sum and difference between the pump and cavity fields wavevectors. Starting from the light-matter interaction Hamiltonian in the dispersive regime and following a canonical mean-field procedure, we derive explicitly the free energy as a function of the atomic ($\Theta,\Pi$) and photonic ($X$) order parameters. 
To analyze the onset of the superradiant transition, we compute the Hessian of the free energy as a function of the coupled order parameters, deriving the analogues of the energy costs $\epsilon_{\Theta/\Pi}$ and coupling $\rm{U}$ from microscopics through a set of corresponding linearized response susceptibilities. 
Within this analysis, the superradiant phase boundary is determined by:
\begin{equation}
    {N \mathcal{D}_{0{\rm C}}} = \frac{8}{\chi_{{\rm n},{\rm n}} +  \left(\frac{ \Tilde{\Omega}_{\mathrm{m}}}{ \Omega_{\mathrm{a}}} \frac{\Delta}{E_{\rm F}} \right) (\chi_{\rm n,\eta}' + \chi_{\rm \eta,n}') + \left( \frac{\Tilde{\Omega}_{\mathrm{m}}}{\Omega_{\mathrm{a}}} \frac{\Delta}{E_{\rm F}} \right)^2 \chi_{\rm \eta,\eta}'},    \label{eqn:main-equation}
\end{equation}

%
where $\mathcal{D}_{0\rm{C}} = \Omega_a^2/\Delta_c$ is the critical reduced light-matter coupling strength, $\Omega_a$ is the dispersive shift per atom, $\Delta$ is the superfluid pairing gap and $\Tilde{\Omega}_\mathrm{m}$ is an effective molecular coupling strength proportional to $1/\Delta_m$. 
The susceptibilities---$\chi_{\rm n,n}$,  $\chi_{\rm n, \eta}'$ and $\chi_{\rm \eta, \eta}'$---characterize fermionic density-density, density-pair-density, and pair-density-pair-density responses at the relevant momenta; for full details on the derivation starting from the microscopic Hamiltonian, see Methods and Ref. \cite{doi:10.1142/q0409}.
The Fano-shaped phase boundary controlled by $r$ in Eq. \eqref{eq:freeEnergy} results from the sign change of the effective molecular coupling $\Tilde \Omega_m$ across the PA resonance, tuning between competing and cooperating orders. Importantly, the width of this resonance profile reflects the strength of the cross-coupling of the two orders $\Theta,\Pi$, which is captured by finite cross-susceptibilities $\Delta \chi_{\rm n, \eta}'  = \Delta \chi_{\rm \eta,n}'$ arising naturally from the strong correlations in the unitary Fermi gas.  

We quantitatively compare the phase boundary given by Eq.~\eqref{eqn:main-equation} to the trap-averaged response function at finite momenta using a generalized random-phase approximation (RPA) to estimate the susceptibilities from first principles. The RPA results are shown by the solid lines in Fig~\ref{fig:fig3}b and are in agreement with the experimental data. We also apply this approach away from unitarity, predicting a reduction of the cross-coupling as the system transitions from BEC to the BCS regime, as seen in the inset of Fig~\ref{fig:fig3}b. In the BEC regime the atom pairs are bound more tightly than in the BCS regime, increasing both the background value of $\Delta$ and the susceptibility $\Delta \chi_{\rm n, \eta}$ coupling the atomic density and the pair-density waves.

The explicit appearance of the pairing gap in the phase boundary equation directly results from the coupling of light to atom pairs in the PA process, and highlights the pair-density wave nature of the ordered phase. More rigorously, light couples to atom pairs at a distance given by the Condon radius, much shorter than the Fermi wavelength. In the above expression, $\Delta$ should therefore be understood as the short-distance pairing field, comprising both condensed and non-condensed pairs \cite{altman:2005aa, Haussmann:2009ab}, and accurately described by Tan's contact \cite{Braaten:2008aa}. Therefore, the PA transition acts as an optical Feshbach resonance \cite{Theis:2004aa,Chin_Grimm_TiesingaFRIUG}, and the pair-density wave can be interpreted as a spontaneous modulation of the contact (see Methods for details). In the long-wavelength approximation where $|\mathbf{Q}| \longrightarrow 0$, this can be used to estimate the susceptibilities from the known variations of the contact with scattering length \cite{Navon:2010ab, Horikoshi:2017aa,jager:2024aa}. The results are shown with dashed lines in Fig. \ref{fig:fig3}, showing a good agreement with the data.


\section{Discussion}


Our results show that strong coupling to light is a new mechanism for the formation of pair-density waves in quantum gases, i.e. a non-trivial spontaneous modulation of the pair-density in a Fermi superfluid, originating from strong interactions. This differs with pair-density waves investigated in the context of strongly correlated electrons \cite{Agterberg:2020aa}: First, it takes the form of a modulation of the short-range pair correlations, usually captured by Tan's contact, and we expect that at high temperature or in the far BCS regime, the distinction between the pairing gap and the contact will require a description beyond our simple mean-field approach. Second, the modulation occurs on top of a large, uniform order parameter background originating from strong contact attraction, rather than as a sign-alternating pairing gap, as for example in the Fulde--Ferrell--Larkin--Ovchinnikov phase \cite{Kinnunen:2018vc,Agterberg:2020aa}. An important open question, both from the theoretical and experimental point of view, is the relationship between our observations of charge and pair order and superfluidity. More generally, the complete phase diagram and the type of transitions occuring in a system of fermions with strong contact interactions together with light-matter coupling to pairs remains to be explored \cite{sharma:2024aa,chen:2025aa}. 

Experimentally, losses at the molecular transition have limited our investigations to situations in which the contact contribution remains smaller than the background atomic one. Losses will likely limit also the lifetime of the phases of matter reached above the threshold, restricting the range of techniques available to probe the nature of the organized phase. Two-electron atoms in optical cavities \cite{norcia:2016ac,kawasaki:2019aa,rivero:2022aa}, for which long-lived PA transitions are known to exist \cite{enomoto:2008aa,blatt:2011aa,hofer:2015aa,pagano:2015ab}, would be particularly suited to further investigate exotic quantum phases that could emerge in situations where pair coupling is dominant. Conversely, two-body, molecular losses are known to give rise to a variety of correlation phenomena \cite{Tomita:2017aa,yamamoto:2021aa,huang:2023ab} that, together with the dissipation induced by the cavity, could be further studied in our experiment.

Last, adiabatically eliminating the cavity in the large detuning regime allows to interpret our system as having an infinite range photon-mediated atom-atom, atom-pair and pair-pair interactions. It demonstrates that cavity QED methods are suited to synthesize strong interactions beyond two-body \cite{luo:2024aa} in previously not accessible regimes \cite{kraemer:2006aa,petrov:2014ac,hammond:2022ab}.

\section*{ACKNOWLEDGEMENTS}

We thank Felix Werner, Markus Mueller, Aurélien Fabre and Gaia Bolognini for useful discussions. 
The EPFL group acknowledges funding from the Swiss State Secretariat for Education, Research and Innovation (Grants No. MB22.00063 and UeM019-5.1). The ETH group acknowledges funding from the SNSF project 200021\_212899, the Swiss State Secretariat for Education, Research and Innovation (SERI) under contract number UeM019-1, and NCCR SPIN, a National Centre of Competence in Research, funded by the Swiss National Science Foundation (grant number 225153).
SU acknowledges funding from JST PRESTO (Grant No. JPMJPR235) and JSPS KAKENHI (Grant No. JP21K03436).


\section*{METHODS}

\subsection*{Experimental procedure}
We start with a degenerate Fermi gas of temperature $T/T_{\mathrm{F}} \approx 0.12$ with $N = 5.2(3) \times 10^5$ ${}^6$Li atoms equally populating the two lowest hyperfine states at unitarity of the broad Feshbach resonance at $832~\mathrm{G}$. The atoms are harmonically trapped with a radial trap frequency of $430~\mathrm{Hz}$ and axial trap frequency of $28~\mathrm{Hz}$ in a hybrid optical and magnetic trap. We induce cavity-mediated long-range interactions by illuminating the cloud from the side using a retro-reflected pump beam with $\pi$-polarization. The pump and the cavity resonance are detuned with respect to the atomic D2 transition by $ -25.25 \mathrm{GHz}$. There, the atoms induce a mean dispersive shift of the cavity resonance by $\delta_{\mathrm{c}} = \Omega_{\mathrm{a}} N/2 = -2\pi \times 2.82(2)~\mathrm{MHz}$ due to the coupling to single atoms, largely exceeding the cavity linewidth $\kappa = 2\pi \times 77(1) ~\mathrm{kHz}$. 

The experiment is performed in the vicinity of a strongly-coupled photoassociation transition, which is located at a detuning of $ -25.247\mathrm{GHz}$ from the zero field $D2$ line of $^6$Li, corresponding to the $\nu = 81$ molecular bound states in the $1\Sigma^+_g$ excited potential. This photoassociation line was already investigated in \cite{KonishiUPPIASIFG2021}, where a single photon-pair coupling strength of $g_{\mathrm{m}} = 2\pi \times 383(3)~\mathrm{kHz}$ was found using a Condon radius $R_{\mathrm{c}} = 164a_0$ and a width of the radial molecular orbital $L = 12.6a_0$. 

We perform linear ramps of the pump strength $V_0 = \Omega_\mathrm{a}|\alpha|^2$ (see below for notations) across the self-organization phase transition, while simultaneously recording the photon flux $\bar{n}_\mathrm{det}$ leaking from the cavity. The cavity photon leakage is detected with an efficiency of $\sim 3\%$ \cite{HelsonOROASIFG2022}. The linear pump ramp allows us to directly convert time into $V_0$. We acquire phase diagrams for detunings from the dispersively-shifted cavity $\Tilde{\Delta}_\mathrm{c} = \Delta_\mathrm{c} - \delta_\mathrm{c}$ between $-0.2~\mathrm{MHz}$ and $-4~\mathrm{MHz}$. The pump lattice depth is calibrated using Kapitza-Dirac diffraction on a molecular BEC far away from the photoassociation transition, directly obtaining $V_{0}$ without a pair-coupling contribution \cite{GadwayAOKDDPBRNR2009}.

The retro-reflected pump beam forms a standing wave that intersects the cavity axis at an angle of $18^\circ$ \cite{HelsonDWOIAUFGWPMI2023}, as presented schematically in Fig.~\ref{fig:fig2}b. This configuration results in different recoil momenta upon Rayleigh scattering between pump and cavity, $\mathbf{k}_\pm = \mathbf{k}_\mathrm{c} \pm \mathbf{k}_\mathrm{p}$. The low-energy mode with momentum $\hbar \mathbf{k}_-$ dominate ordering at the phase boundary \cite{ZwettlerNEDOLRIF2024}. 

\subsection*{Data analysis}
\begin{figure}[h!]
\includegraphics{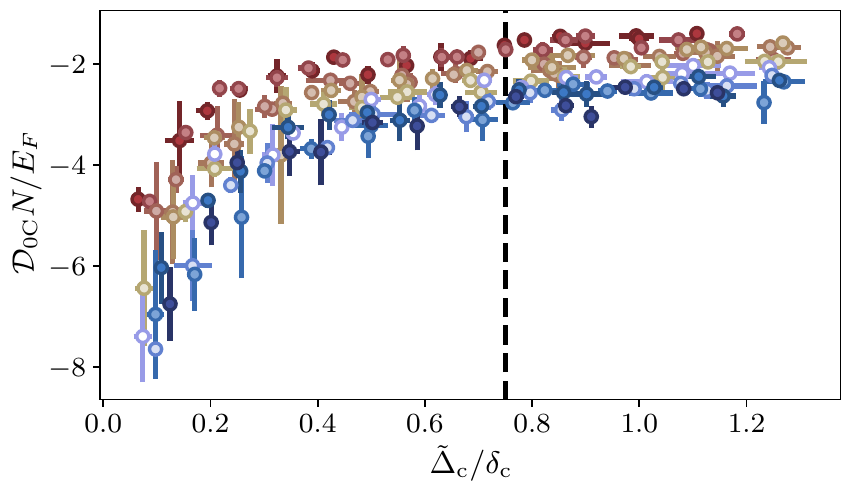}
\caption{\textbf{Critical Long-Range Interaction Strength.} $\mathcal{D}_{0 \mathrm{C}}N/E_{\mathrm{F}}$ at unitarity as a function of the normalized dispersively-shifted cavity detuning $\Tilde{\Delta}_\mathrm{c}/\delta_\mathrm{c}$ for various molecular detunings $\Delta_\mathrm{m}$: $-100~\mathrm{MHz}$ ($\textcolor{bilbao9_2}{\circ}$), $-125~\mathrm{MHz}$ ($\textcolor{bilbao9_3}{\circ}$), $-175~\mathrm{MHz}$ ($\textcolor{batlow9_4}{\circ}$), $-200~\mathrm{MHz}$ ($\textcolor{bilbao9_5}{\circ}$), $-300~\mathrm{MHz}$ ($\textcolor{bilbao9_6}{\circ}$), $-400~\mathrm{MHz}$ ($\textcolor{bilbao9_7}{\circ}$), $400~\mathrm{MHz}$ ($\textcolor{devon9_6}{\circ}$), $300~\mathrm{MHz}$ ($\textcolor{devon9_5}{\circ}$), $200~\mathrm{MHz}$ ($\textcolor{devon9_4}{\circ}$), $125~\mathrm{MHz}$ ($\textcolor{devon9_3}{\circ}$), $100~\mathrm{MHz}$ ($\textcolor{devon9_2}{\circ}$).}
\label{fig:sup1}
\end{figure}

We fit the onset of density wave ordering as a function of $V_0$ i.e. the critical pump strength on each experimental trace using a linear function:

\begin{equation}
    \bar{n}_{\mathrm{det}}(V_0) = \theta(V_0 - V_{0\mathrm{C}})\times B(V_0 - V_{0\mathrm{C}}),
\end{equation}

with $\theta(V_0 - V_{0\mathrm{C}})$ the Heaviside function and fit parameters for critical pump strength $V_{0\mathrm{C}}$ and the slope of the photon flux onset $B$. Each extracted $V_{0\mathrm{C}}(\Tilde{\Delta}_\mathrm{c})$ is subsequently converted in a long-range interaction strength $\mathcal{D}_{0 \mathrm{C}}(\Tilde{\Delta}_\mathrm{c}) = \Omega_{\mathrm{a}}V_{0\mathrm{C}}(\Tilde{\Delta}_\mathrm{c})/\Tilde{\Delta}_\mathrm{c}$ using the atomic dispersive coupling strength $\Omega_{\mathrm{a}}$, as can be seen from Fig.~\ref{fig:sup1}. To render $\mathcal{D}_{0 \mathrm{C}}$ dimensionless, we use the total atom number $N$ and the corresponding Fermi energy in the harmonic trap $E_{\mathrm{F}}$ from absorption imaging, which is acquired together with the data for density wave ordering in a randomized manner. We observe a systematic shift of $\mathcal{D}_{0 \mathrm{C}}N/E_{\mathrm{F}}$ to higher absolute values for small $\Tilde{\Delta}_\mathrm{c}$, which is due to the finite ramp speed of $\dot{V}_0 = 1.75 \times E_\mathrm{R}/\mathrm{ms}$. Therefore, we extract $\mathcal{D}_{0 \mathrm{C}}$ by averaging values at large detunings of $\Tilde{\Delta}_\mathrm{c}/\delta_\mathrm{c} > 0.75$. We still observe an absolute atomic value at unitarity of $D_{0\mathrm{C, a}}N/E_\mathrm{F} = -2.05$, which is a factor of two higher than the value reported in \cite{ZwettlerNEDOLRIF2024}.

\subsection*{Atom Losses}

\begin{figure}[h!]
\includegraphics{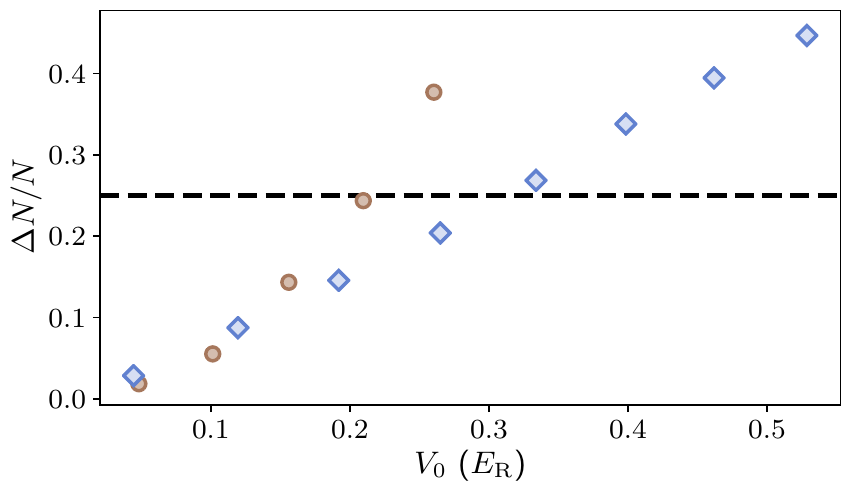}
\caption{\textbf{Atom Loss during $V_0$ Ramp.} Atom Loss $\Delta N /N$ at unitarity at $\Delta_{\rm{m}}=-100~\mathrm{MHz}$ ($\textcolor{bilbao9_5}{\medcircle}$) and $+100~\mathrm{MHz}$ ($ \textcolor{devon9_5}{\meddiamond}$) during a linear ramp of $V_0$ measured for various endpoints while keeping the ramp speed fixed. The uncertainties are smaller than the marker size and thus not discernible in the plot.}
\label{fig:sup2}
\end{figure}

We estimate the upper bound on the atom loss during the linear ramp of $V_0$ until the critical pump strength $V_{0\mathrm{C}}$ is reached, by measuring the losses at molecular detunings of $\Delta_{m} = \pm 100~\mathrm{MHz}$ using two consecutive dispersive shift measurements before and after the $V_0$ ramp with a positive pump-cavity detuning, preventing the atoms to undergo self-organization. The results are shown in Fig. \ref{fig:sup2}. This yields a maximal atom loss of $\sim25\%$ until $V_{0C}$ is reached, which is encountered at the largest pump-cavity detuning $\Tilde{\Delta}_c$. We use $\sigma$-polarization for the loss measurement with the dispersive shift to avoid the systematic dispersive shift at $\pi$-polarization due to coupling to the photoassociation transition, which would make the apparent losses smaller (larger) for positive (negative) detunings.

\subsection*{Dispersively-Coupled Light-Matter Hamiltonian}

In this section, we derive the Hamiltonian of the transversely pumped atom-cavity system with dispersive coupling to atoms and pairs. The Fermi gas inside the optical resonator is illuminated by a standing-wave, retro-reflected pump beam with a pump lattice amplitude $\alpha$ and geometry as described in \cite{HelsonDWOIAUFGWPMI2023}. Due to the close detuning to a photoassociation transition, the combined pump and cavity light field $\hat{\phi}$ is not only coupled to atomic density, but also to a pair density. This results in a new type of dispersive light-matter interaction Hamiltonian:

\begin{equation}
    \hat{H}_{\mathrm{int}} = \int d^3\mathbf{R}\left(\Omega_{\mathrm{a}} \hat{n}(\mathbf{R}) + \Omega_{\mathrm{m}} \hat{\mathcal B}(\mathbf{R})\right)\hat{\phi}^{\dagger}(\mathbf{R}) \hat{\phi}(\mathbf{R})
\end{equation}

with dispersive light-matter coupling strengths ${\Omega}_{i} = g^2_{i}/ \Delta_{i}$ for single atoms and pairs, $i = a, m$, the atomic density operator $\hat{n}(\mathbf{R})$ and the pair density operator $\hat{\mathcal B}(\mathbf{R}) = \hat{P}^{\dagger}(\mathbf{R})\hat{P}(\mathbf{R})$. The pair annihilation operator is given by:

\begin{equation}
    \hat{P}(\mathbf{R}) = \int d\mathbf{r}f(\mathbf{r})\hat{\psi}_{\uparrow}(\mathbf{R}-\frac{\mathbf{r}}{2})\hat{\psi}_{\downarrow}(\mathbf{R}+\frac{\mathbf{r}}{2})
\end{equation}

with $f(\mathbf{r})$ the molecular orbital of an excited molecular state to which the ground-state pair is coupled. The intensity of the total light field is:
\begin{align}
    \hat{\phi}^{\dagger}(\mathbf{R})\hat{\phi}(\mathbf{R}) & = \cos^2(\mathbf{k}_\mathrm{c}\cdot\mathbf{R})\hat{a}^{\dagger} \hat{a} + \cos^2(\mathbf{k}_\mathrm{c}\cdot\mathbf{R})\alpha^2 \\
    & + \cos(\mathbf{k}_\mathrm{c}\cdot \mathbf{R})\cos(\mathbf{k}_\mathrm{p}\cdot \mathbf{R}) \alpha (\hat{a}^{\dagger} + \hat{a}), \nonumber
\end{align}

where we have chosen a real pump field amplitude $\alpha = \alpha^*$, and $\hat{a}$ is the quantized cavity field. 
In the following, we assume $\Omega_\mathrm{a}$ to be constant, given the small variations of detuning to atomic transition. We approximate the molecular orbital $f(\mathbf{r})$ as  a box centered at a Franck-Condon radius $R_c$ with a width $L$ \cite{KonishiUPPIASIFG2021}:

\begin{equation}
        f(r) =  \left\{
        \begin{array}{ll}
           \frac{1}{\sqrt{4 \pi L}} \frac{1}{r}  & \mbox{if } R_\mathrm{c}-\frac{L}{2}<r<R_\mathrm{c}+\frac{L}{2}\\
            0 & \mbox{otherwise}
        \end{array}\right.
        \label{eqn:orbital}
\end{equation}

In the following, we neglect the pump-lattice potential for atoms and pairs. By introducing the dispersive shift contributions due to coupling to atoms and pairs:

\begin{align}
    {\delta}_{\mathrm{c,a}} &= \frac{g_{\mathrm{a}}^2}{\Delta_{\mathrm{a}}}\int d\mathbf{R} {n}(\mathbf{R})\cos^2(\mathbf{k}_\mathrm{c}\cdot\mathbf{R}) \\
    {\delta}_{\mathrm{c,p}} &=  \frac{g_{\mathrm{m}}^2}{\Delta_{\mathrm{m}}}\int d\mathbf{R} {\mathcal{B}}(\mathbf{R})\cos^2(\mathbf{k}_\mathrm{c}\cdot\mathbf{R}),
\end{align}

we arrive at the full dispersively-coupled light-matter Hamiltonian:

\begin{equation}
    \begin{split}
    \hat{H}_{\mathrm{lm}}  =\hat{H}_{\mathrm{at}} -& \Tilde{\Delta}_{\mathrm{c}} \hat{a}^{\dagger}\hat{a} + \frac{1}{4} \alpha \sum_{\mathbf{Q}}(\hat{a}+\hat{a}^{\dagger}) \\ 
    & \times \int d\mathbf{R} \left(\Omega_{\mathrm{a}} \hat{n}(\mathbf{R}) + \Omega_{\mathrm{m}} \hat{\mathcal{B}}(\mathbf{R}) \right) e^{i \mathbf{Q}  \cdot \mathbf{R} }
    \label{eqn:effective-hamiltonian}
    \end{split}
\end{equation}

with the Hamiltonian for a trapped, interacting Fermi gas $\hat{H}_{\mathrm{at}}$, $ \mathbf{Q} = \pm (\mathbf{k}_p \pm \mathbf{k}_c)$ and the dispersively-shifted cavity detuning $\tilde{\Delta}_{\mathrm{c}} = (\Delta_{\mathrm{c}} - {\delta}_{c, a} - {\delta}_{c, p})$.

\subsection*{Mean-field free energy} \label{sec:mf-energy}

In this section, we derive the effective free energy of the system as in Eq. ~\eqref{eq:freeEnergy}, within the mean-field approximation. The bare atomic Hamiltonian $H_{\rm at}$ is given by the mean-field BCS Hamiltonian 

\begin{equation}
    H_0 = \sum_{k,\sigma} (\xi_k - \mu) \hat{c}^\dag_{k,\sigma} \hat{c}_{k,\sigma} + \Delta \sum_k \hat{c}^\dag_{k,\uparrow} \hat{c}^\dag_{k,\downarrow} + \text{H.c.}, 
\end{equation}

where $\Delta$ is the BCS gap parameter, $\mu$ is the chemical potential, $\xi_k = \frac{k^2}{2m}$ is the free fermion dispersion and $m$ is the atomic mass. To derive the effective free energy of the system, we mean-field decouple the interaction in the light-matter Hamiltonian given by Eq.~\eqref{eqn:effective-hamiltonian}. 
We expand the Hamiltonian, keeping all terms up to quadratic order in cavity field $x$, density operator  $\hat{\theta}_\mathbf{Q} = \sum_{k,\sigma} \hat{c}^\dag_{k+Q,\sigma}\hat{c}_{k,\sigma}$ and pair-density operator $\hat{\eta}_{k,\mathbf{Q}} = \hat{c}_{-k + Q,\downarrow} \hat{c}_{k}$. The light-matter coupling can then be grouped into two parts, coupling to the fermionic density 

\begin{equation}
     H_{\mathrm{int}, \rho} = \left(\Omega_{\rm a} + \Omega_{\rm m} \frac{4\pi}{3}k_F^3 |f_0|^2 \right)  \frac{\alpha}{2\sqrt{2}} \hat{x} \sum_Q \hat{\theta}_Q,
\end{equation}

or to the fermionic pair-density 

\begin{equation}
\begin{split}
       H_{\mathrm{int}, \eta} &=\Omega_{\rm m} \frac{\alpha}{2\sqrt{2}} \hat{x} |f_0|^2 \frac{(2\pi)^3}{V}  \times \\
       &\sum_{k_1,k_2,Q}^{k_M}\left(\langle\hat{\eta}_{k_1, 0}^\dag\rangle \hat{\eta}_{k_2, Q} + \langle \hat{\eta}_{k_1,0}\rangle \hat{\eta}_{k_2,Q}^\dag\right),
\end{split}
\end{equation}

where $\hat{x} = \frac{1}{\sqrt{2}}(\hat{a} + \hat{a}^\dag)$ and $f(\mathbf{k})$ is the Fourier transform of the molecular wave function $f(\mathbf{r})$ given in Eq.~\eqref{eqn:orbital}, which to first approximation simply provides a cut-off in momentum at $k_M = 1/R_c$ and a total molecular volume factor $f_0 = \frac{1}{\sqrt{2}\pi}\sqrt{R_C^2 L}$. Expanding the Hamiltonian in powers of Fermi-wavevector and the Frank-Condon Radius $k_{\rm F} R_c \ll 1$, simplifies to total interaction term to take the form
\begin{equation}
    H_{\mathrm{int}} = \Omega_{\rm a} \frac{\alpha}{2\sqrt{2}} \hat{x} \sum_\mathbf{Q} \hat{\theta}_\mathbf{Q}  + \Tilde{\Omega}_m \frac{\Delta}{E_{\rm F}} \frac{\alpha}{2\sqrt{2}} \hat{x} \sum_\mathbf{Q} (\hat{\eta}'_{\mathbf{Q}} + {\hat{\eta}'^{\dag}}_\mathbf{Q}),
\end{equation} 

where $\hat{\eta}' =  \frac{8 k_{\rm F} R_C }{3\pi^2} \sum_k^{k_M} \hat{\eta}_{k,\mathbf{Q}}$ an integrated pairing operator, with the cut-off set by the molecular wavefunction and $\Tilde{\Omega}_m = \Omega_m \frac{3 \pi k_F L}{16\pi} $. Note, in all the computations we keep track of the molecular cut-off and assume the hierarchy of scales $k_{\rm F} \ll \frac{1}{R_c} \ll \frac{1}{b}$, where $b$ is the contact interaction range. 
Having derived the full mean-field Hamiltonian, we now proceed to compute the effective free energy in terms of the order parameters. We introduce three source terms $h_x,h_\Theta,h_\eta$ , coupling to the order parameters and  write the partition function as

\begin{equation}
    Z[h] = \Tr\left[ e^{-\beta\left(\hat{H}_{\rm lm} - h_X\hat{x} - \frac{1}{4}\sum_{\mathbf{Q}} h_\Theta \hat{\theta}_\mathbf{Q} +  h_\Pi \hat{\eta}'_\mathbf{Q} \right)} \right]
\end{equation}

The order parameters are defined as the derivatives of the partition function

\begin{align}
    \Theta &=  \frac{\partial \log Z}{\partial (\beta h_\Theta)} = \left\langle\frac{1}{4}\sum_{\mathbf{Q}} \hat{\theta}_\mathbf{Q} \right\rangle \\
\Pi &=  \frac{\partial \log Z}{\partial (\beta h_\Pi)} = \left\langle\frac{1}{4}\sum_{\mathbf{Q}} \hat{\eta}'_\mathbf{Q} + \hat{\eta}'^\dag_{\mathbf{Q}} \right\rangle\\
X &=  \frac{\partial \log Z}{\partial (\beta h_X)}  = \left\langle\hat{x}\right\rangle.
\end{align}

To obtain the Landau-Ginzburg free energy, we perform a Legendre transform with respect to the source terms

\begin{equation}
    \mathcal{F}[\Theta,X,\Pi] = -\frac{1}{\beta} \Tr[\log Z[h]] + h_\Theta \Theta + h_\Pi \Pi + h_X X  
\end{equation}

Following \cite{georges:2004aa, doi:10.1142/q0409}, we consider the free energy as parametrically dependent on the pump strength $\alpha$. The Hellman-Feynman theorem imposes 
\begin{equation}
    \frac{\partial\mathcal{F}}{\partial\alpha} = \left\langle \frac{\partial \hat{H}_{\rm lm}}{\partial\alpha}\right\rangle = \left\langle \frac{1}{2\sqrt{2}} \hat{x} \sum_{\mathbf{Q}} \Omega_{\rm a} \hat{\theta}_\mathbf{Q} + \frac{\Delta}{E_{\rm F}}\Tilde{\Omega}_{\rm m} (\hat{\eta}'_{\mathbf{Q}} + \hat{\eta}'^\dag_\mathbf{Q}) \right\rangle_{\alpha}
\end{equation}

where the expectation value is evaluated with a finite value of $\alpha$. Integrating from $\alpha=0$, we obtain the (so far exact) expression:
\begin{multline}
\mathcal{F}(X,\Pi,\Theta) = \mathcal{F}^{(0)}(X,\Pi,\Theta) + \\ \int_0^\alpha d\alpha' \left\langle \frac{1}{4} (\hat{a} + \hat{a}^\dag) \sum_{\mathbf{Q}} \Omega_{\rm a} \hat{\theta}_\mathbf{Q} +\frac{\Delta}{E_{\rm F}} \Tilde{\Omega}_{\rm m} (\hat{\eta}'_{\mathbf{Q}} + \hat{\eta}'^\dag_\mathbf{Q})  \right\rangle_{\alpha'}
\end{multline}

where the expectation value is taken for the many-body state of the system at the running value of the pump strength $\alpha'$. $\mathcal{F}^{(0)}$ is the free energy of the atoms and cavity field in the absence of coupling via pump laser. Up to linear order in the coupling between the fermions and the cavity field, we can re-write the Free energy as

\begin{equation}
\mathcal{F}(X,\Pi,\Theta) =  \mathcal{F}^{(0)}(X,\Pi,\Theta) - \Lambda X (\Theta + r \Pi),
\end{equation}
where we have ignored the dependence of the expectation value on $\alpha'$ and separated the fermionic and cavity degrees of freedom in the expectation value, as in the absence of external coupling the expectation value decouples.  We also identified $\Lambda= - \sqrt{2} \Omega_a$ and $r= \frac{\bar{\Omega}_m}{\Omega_a} \frac{\Delta}{E_{\rm F}}$. The uncoupled part of free energy can be split into cavity and fermionic part $\mathcal{F}^{(0)} = \Delta_c X^2 + \mathcal{F}^{(F)}$, where $\mathcal{F}^{(F)}$ is the free energy of interacting free fermions. For the fermionic variables $\Theta, \Pi$, by performing the Legendre transformation as defined above it follows that 
\begin{equation}
    \frac{\delta \mathcal{F}^{(F)}[\Pi,\Theta]}{\delta \Theta} = \beta h_{\Theta} \quad  \frac{\delta \mathcal{F}^{(F)}[\Pi,\Theta]}{\delta \Pi} = \beta h_{\Pi}
\end{equation}

To obtain the free energy up to the quadratic coefficient, we need to compute the second derivative of the free energy, w.r.t. the order paramereters i.e. $\frac{\delta^2 \mathcal{F}^{(F)}}{\delta \Theta \delta\Theta}$. By differentiating the above equation we obtain:
\begin{equation}
    \sum_{k} ( \delta_i \delta_k \mathcal{F}^0 ) \times \frac{\delta^2 \log Z[h]}{\delta (\beta h_k) \delta (\beta h_j)} = \delta_{i,j}  
\end{equation}

By definition of the response functions, it follows $\frac{1}{N} \frac{\delta^2 \log Z[h]}{\delta (\beta h_k) \delta (\beta h_j) } = \chi_{k,j}$, where $N$ is the fermion number. Inverting the above relation gives

\begin{equation}
    \begin{split}
    \mathcal{F}^{(F)}[\Theta,\Pi] = \frac{1}{2N} &\left( \frac{\chi'_{\rm \eta,\eta}}{\chi_{\rm n,n}\chi'_{\rm \eta,\eta} - |\chi'_{\rm \eta,n}|^2}\right) \Theta^2\\
    + &\frac{1}{2N}\left(\frac{\chi_{\rm n,n}}{\chi_{\rm n,n}\chi_{\rm \eta,\eta}' - |\chi_{\rm \eta,n}'|^2}\right) \Pi^2 \\
    &- \frac{1}{N} \left(\frac{\Re{\chi_{\rm \eta,n}'}}{\chi_{\rm n,n}\chi_{\rm \eta,\eta}' - |\chi'_{\rm \eta,n}|^2}\right) \Theta\Pi 
    \end{split}
\end{equation}

where $\chi_{\rm n,n}$ is the density-density response function of the system, $\chi_{\rm n,\eta}'$ is the regularized density-pairing response and $\chi_{\rm \eta,\eta}'$ is the regularized pairing-pairing response function. The response functions are evaluated using the RPA response~\cite{ZhaoDSFOATDFSWRPA2020, MarijanovicDIOSIUFIAOC2024}, keeping track of the molecular cut-off $k_M$. We can now identify $\epsilon_{\Theta} = \frac{1}{2N} \left( \frac{\chi'_{\rm \eta,\eta}}{\chi_{\rm n,n}\chi_{\rm \eta,\eta}' - |\chi_{\rm \eta,n}'|^2}\right)  $, $\epsilon_\Pi = \frac{1}{2N} \left( \frac{\chi_{n,n}}{\chi_{\rm n,n}\chi_{\rm \eta,\eta}' - |\chi_{\rm \eta,n}'|^2}\right) $ and $U = \frac{1}{N}\left(\frac{\Re{\chi_{\rm \eta,n}'}}{\chi_{\rm n,n}\chi_{\rm \eta,\eta}' - |\chi_{\rm \eta,n}'|}\right)$.

\subsection*{Phase-boundary}

To compute the phase boundary, we compute the Hessian of the Free energy and require that one of the normal modes goes soft, i.e. the determinant of the Hessian is zero. Using notation from the main text, this gives the boundary equation 

\begin{equation}
    \frac{\Lambda_c^2}{\Delta_c} = \frac{2\epsilon_\Theta \epsilon_\Pi - U^2/2}{\epsilon_\Pi + rU + r^2 \epsilon_\Theta}.
    \label{eqn:crit}
\end{equation}

 The equation in terms of the microscopic response functions reduces to

\begin{equation}
    {N \mathcal{D}_{0{\rm C}}} = \frac{8}{\chi_{{\rm n},{\rm n}} +  \left(\frac{ \Tilde{\Omega}_{\mathrm{m}}}{ \Omega_{\mathrm{a}}} \frac{\Delta}{E_{\rm F}} \right) (\chi_{\rm n,\eta}' + \chi_{\rm \eta,n}') + \left( \frac{\Tilde{\Omega}_{\mathrm{m}}}{\Omega_{\mathrm{a}}} \frac{\Delta}{E_{\rm F}} \right)^2 \chi_{\rm \eta,\eta}'},    \label{eqn:main-equation}
\end{equation}

where $\mathcal{D}_{0{\rm C}} = \Omega_a^2/\Delta_a$ is the effective pump power.

\subsection*{Pair-density wave and Tan's contact}

After deriving the boundary equation in the mean-field picture, we generalize the approach to take into account the full many-body correlations. Our approach consists of two main steps: first, evaluating the molecular density operator $\hat{\mathcal{B}}(\mathbf{R})$, and second, employing linearized equations of motion, generalizing the previously developed approach in \cite{MarijanovicDIOSIUFIAOC2024, ZwettlerNEDOLRIF2024}. Given the separation of scales, $b \ll R_c \ll 1/k_{\rm F}$, where $b$ is the range of the short-range contact interactions from the Feshbach resonance, we can utilize the operator product expansion \cite{ZwergerTBCS-BECCATUFG2012, ProcEnricoFermi2016} to obtain:

\begin{equation}
   \hat{\mathcal{B}}(\mathbf{R})  = \hat{C}(\mathbf{R}) \left| \int d\mathbf{r} f(\mathbf{r}) \frac{1}{4\pi|\mathbf{r}|}\right|^2
\end{equation}

up to leading order in $1/|\mathbf{r}|$. We define the operator $\hat{C}(\mathbf{R}) = m^2U^2 \hat{\psi}^\dag_{\uparrow}(\mathbf{R})\hat{\psi}^\dag_{\downarrow}(\mathbf{R}) \hat{\psi}_{\downarrow}(\mathbf{R})\hat{\psi}_{\uparrow}(\mathbf{R}) $, 
which upon taking the expectation value is contact density at position $\mathbf{R}$, $\langle\hat{C}(\mathbf{R})\rangle = C(\mathbf{R})$, and $U$ represents the contact interaction strength, requiring appropriate renormalization (see below). Using the simple excited-state molecular wave function model given in Eq.~\eqref{eqn:orbital}, we obtain for the ground state and excited state overlap integral:

\begin{equation}
    \left| \int d\mathbf{r} f(\mathbf{r}) \frac{1}{4\pi |\mathbf{r}|}\right|^2 = \frac{L}{4\pi}.
\end{equation}

Absorbing the molecular factor into the coupling constant gives the final effective Hamiltonian

\begin{equation}
    \hat{H}_{\rm lm} = \hat{H}_{at} - \Tilde{\Delta}_{\mathrm{c}} \hat{a}^\dag \hat{a} + \frac{1}{4} \alpha (\hat{a} + \hat{a}^\dag) \sum_{\mathbf{Q}} \Omega_{\rm a} \hat{\theta}_\mathbf{Q} + \Tilde{\Omega}_{\rm m} \hat{\Pi}_\mathbf{Q}  
    \label{eqn:Hamiltonian-drive}
\end{equation}

where now we re-define $\Tilde{\Omega}_{\mathrm{m}} = k_{\rm F} L \Omega_{\mathrm{m}}/4\pi $, $k_{\rm F}$ is the Fermi wave vector and the density and contact-density wave operator at wave vector $\mathbf{Q}$ are now defined as

\begin{align}
    \hat{\theta}_\mathbf{Q} &= \int d \mathbf{R} \hat{n}(\mathbf{R}) e^{i \mathbf{Q} \cdot \mathbf{R}} \\
    \hat{\Pi}_\mathbf{Q} &= \int d \mathbf{R} \frac{\hat{C}(\mathbf{R})}{k_F} e^{i \mathbf{Q} \cdot \mathbf{R}}.
\end{align}

Therefore, we see that the light-matter Hamiltonian contains two different ordering channels: the density and the contact-density. Having separated the Hamiltonian in this form, the boundary equation directly follows, with the mean-field response functions replaced $\frac{\Delta}{E_{\rm F}} \chi_{n,\eta}' \rightarrow \chi_{n,C}$ and $\frac{\Delta^2}{E_{\rm F}} \chi_{\eta,\eta}' = \chi_{C,C}$, where $\chi_{n,C}$ is the density-contact and $\chi_{C,C}$ is the contact-contact response function. The boundary equation can be written as
\begin{equation}
    \frac{8}{N \mathcal{D}_{0{\rm C}}}  =  \chi_{{\rm n},{\rm n}} +  \left(\frac{\Tilde{\Omega}_{\mathrm{m}}}{\Omega_{\mathrm{a}}}\right) (\chi_{\rm n,C} + \chi_{\rm C,n}) + \left( \frac{\Tilde{\Omega}_{\mathrm{m}}}{\Omega_{\mathrm{a}}} \right)^2 \chi_{\rm C,C}
\end{equation}

Note, the only approximation in this approach is the decoupling of fermionic and cavity degrees of freedom, which should be valid close to the phase transition. The main challenge remains evaluating the density-contact and the contact-contact response functions. 

\subsection*{Zero-Momentum Response Functions}

In the zero-momentum limit, $\mathbf{Q} \rightarrow 0$, the phase boundary equation can be exactly evaluated using thermodynamic relations. To illustrate the general approach, consider the density-density response, which reduces to the compressibility: $\chi_{n,n} \xrightarrow{\mathbf{Q} \rightarrow 0 }  - \frac{1}{N} \frac{\partial N}{\partial \mu }$, where $N$ is the total atom number. To derive this relation, we introduce an external perturbation coupling to density,  $H_\mathrm{ext} = \phi \theta_Q$. As $Q \rightarrow 0 $, this simplifies to $H_\mathrm{ext} = \phi N$. In the grand-canonical ensemble, we identify $\phi = - \delta \mu$, meaning that adding a small external field corresponds to shifting the chemical potential $\mu$ by $-\delta \mu$. From the standard definition of the response function, it follows that:

\begin{equation}
    \chi_{{\rm n,n}} = \frac{1}{N} \frac{\expval{\theta_Q}}{\phi} \xrightarrow{q \rightarrow 0 }  - \frac{1}{N} \frac{\partial N }{\partial \mu }.
\end{equation}

To derive analogous relations for the case of an external perturbation coupling to the contact, we consider $H_\mathrm{ext} = \phi' \Pi_Q$. Taking the limit $Q \rightarrow 0$ limit, we find that $\expval{\Pi_q} \rightarrow C$, which corresponds to the total contact. By comparing the familiar adiabatic sweep relation $\delta E -= \frac{1}{4\pi m} C \delta(a^{-1})$ \cite{ZwergerTBCS-BECCATUFG2012} with the energy variation due to the external perturbation $\delta E = C \phi'$, we can directly identify $\phi' = \frac{1}{4\pi m} \delta(-a^{-1})$. It is important to note that this identification assumes the chemical potential $\mu$ remains constant as $1/a$ is varied. Following this identification, we obtain:

\begin{align}
    \chi_{{\rm C,n}} &= \frac{1}{N} \frac{\expval{\Pi_Q}}{\phi} \rightarrow \left. - \frac{1}{N}\frac{\partial C}{\partial \mu}\right|_{a} \\ 
    \chi_{{\rm n, C}} &= \frac{1}{N}\frac{\expval{\theta_Q}}{\phi'} \rightarrow \left.  4\pi m  \frac{1}{N}\frac{\partial N}{\partial (-1/a)}\right|_{\mu}\\ 
    \chi_{{\rm C, C}} &= \frac{1}{N}\frac{\expval{\Pi_Q}}{\phi'} \rightarrow \left.  4\pi m  \frac{1}{N} \frac{\partial C}{\partial (-1/a)}\right|_{\mu}.
\end{align}

Using the relations between $N,C,\mu$ in the microcanonical ensemble, these can be related to second derivatives of energy $E(N,a)$

\begin{align}
    \chi_{{\rm n,n}} &=  -\frac{1}{N} \frac{1}{\frac{\partial^2 E}{\partial N^2}}\\
    \chi_{{\rm C,n}} &= -\frac{1}{N} \frac{\partial^2 E}{\partial(-1/a) \partial N} \frac{\partial N}{\partial \mu} 4\pi m \\
    \chi_{{\rm n,C}} &= - \frac{1}{N} \frac{\partial^2 E}{\partial N \partial (-1/a)} \frac{\partial N}{\partial \mu}4\pi m \\
    \chi_{{\rm C,C}} &= \frac{(4\pi m)^2}{N} \frac{\partial^2 E}{\partial (-1/a)^2}  \\
    &+ \frac{(4\pi m)^2}{N}  \left(\frac{\partial^2 E}{\partial N \partial(-1/a)} \right)^2 \left(-\frac{\partial N}
    {\partial \mu}\right),
\end{align}

where we keep $N$, $a$ fixed when evaluating $\partial/\partial a$, $\partial/\partial N$ respectively. The above expression explicitly satisfies the stability condition, $\det(\chi) > 0$, while ensuring that all the response functions remain negative, as required by stability. To evaluate the responses, we use the internal energy expansion at unitarity in terms of $1/k_F a$ at zero temperature \cite{BulgacCOOATFGNTUL2005, ChangQMCSOSFG2004} 

\begin{equation}
    E(N,a) = \frac{3}{5} N \frac{k_{\rm F}^2}{2m} \left( \xi - \frac{\zeta}{k_{\rm F} a} - \frac{5\nu}{3 (k_{\rm F} a)^2 } + \cdots \right),
\end{equation}

where we use $\xi \approx 0.383$, $\zeta \approx 0.901$ and $\nu \approx 0.49$ from Monte Carlo simulations \cite{GandolfiBECBCSCAURIUFG2011}. The resulting phase boundary curve for the unitary Fermi gas is plotted in Fig.~\ref{fig:fig3}b in the main text.

\newpage
\clearpage

\setcounter{figure}{0} 
\renewcommand{\thefigure}{E\arabic{figure}} 
\renewcommand{\figurename}{EXTENDED DATA FIG.}

\end{document}